\documentclass[prb,twocolumn,showpacs,preprintnumbers,amsmath,amssymb,floatfix]{revtex4-1}

\usepackage{color}
\usepackage{graphicx}
\usepackage{longtable}
\usepackage{dcolumn}
\usepackage{bm}
\usepackage{fancybox}

\newcommand{\bfk}{{\bf k}}
\newcommand{\bfr}{{\bf r}}

\newcommand{\req}[1]{\mbox{Eq.~\!(\ref{#1})}}

\def\G0{G^0}

\def\connect#1{\leavevmode{\setbox1=\hbox{#1}\copy1%
\raise .2\ht1 \vbox{\moveleft \wd1\vbox{\hrule width \wd1 height .5pt depth 0pt}}%
}}

\def\ftn[#1]{\rlap{\footnotemark[#1]}}

\def\H0{H^0}
\def\HMZ{H^0_{\sf M}}
\def\hHMZ{\widehat H ^0_{\sf M}}

\def\UM{U_{\sf M}}
\def\hUM{\widehat U_{\sf M}}
\def\UMB{\bar{U}_{\sf M}}
\def\Um{U_{\sf m}}

\def\PM{P_{\sf M}}

\def\WM{W_{\sf M}}

\def\HM{H_{\sf M}}
\def\hHM{\widehat H_{\sf M}}

\def\WFP{W}

\def\Pm{P_{\sf m}}

\def\ww{w}

\makeatletter
\def\Hline{%
\noalign{\ifnum0=`}\fi\hrule \@height 1pt \futurelet
\reserved@a\@xhline}
\makeatother

\begin{document}
\title{Model-mapped random phase approximation to evaluate superconductivity in the 
fluctuation exchange approximation from first principles}

\author{Hirofumi Sakakibara$^{1,2}$}
\email{sakakibara.tottori.u@gmail.com}
\author{Takao Kotani$^{1}$}

\affiliation{$^1$Department of applied mathematics and physics, 
Tottori university, Tottori 680-8552, Japan}
\affiliation{$^2$Computational Condensed
  Matter Physics Laboratory, RIKEN, Wako, Saitama 351-0198, Japan} 

\date{\today}

\begin{abstract}
We have applied
the model-mapped RPA [H. Sakakibara {\it et al.},  J. Phys. Soc. Jpn. {\bf 86}, 044714 (2017)]
to the cuprate superconductors La$_2$CuO$_4$ and HgBa$_2$CuO$_4$,
resulting two-orbital Hubbard models.
All the model parameters are determined based on first-principles calculations.
For the model Hamiltonians, we perform fluctuation exchange
calculation. Results explain relative height of $T_c$ observed in experiment
 for La$_2$CuO$_4$ and HgBa$_2$CuO$_4$.
In addition, we give some analyses for the interaction terms in the model,
especially comparisons with those of the constrained RPA.
 
\end{abstract}
\maketitle


\section{introduction}  It is not so easy to treat strongly-correlated electrons
only by first-principles calculations.
Thus we often use a procedure via a model Hamiltonian \cite{honerkamp,cfRG};
we determine a model Hamiltonian $\hHM$ from a first-principles
calculation and then solve the model Hamiltonian. 
This is inevitable because first-principles calculations, which are
mainly based on the density functional theory (DFT) in the local density approximation
(LDA), are very limited to handle systems with correlated electrons.
Widely used model Hamiltonians are the Hubbard ones,
which consist of one-body Hamiltonian $\hHMZ$ and the on-site
interactions $\hUM$. 
To solve the Hubbard models, we can use a variety of methods 
\cite{DMRG,QMC,VMC,DMFT,TPSC1,TPSC2,onari,tsuchiizu} such as
fluctuation exchange approximation (FLEX) \cite{FLEX}. 

To determine $\hHM$, we have formulated the model-mapped random phase
approximation (mRPA) in Ref. \onlinecite{mRPA} recently.
In mRPA, we use the standard procedure of 
the maximally localized Wannier function \cite{marzari_maximally_1997,souza_maximally_2001}
to determine $\hHMZ$. Here $\hHMZ$ is determined as a projection of the one-body
Hamiltonian of first-principles onto a model space, which is spanned by the Wannier functions.
Then we determine $\hUM$ so that the screened interaction of the model in the random phase approximation
(RPA) agrees with that of the first-principles. 
In this paper, we consider on-site-only interaction in the model.
Then we determine one-body double-counting term $\UMB$. 
Finally we have $\hHM=\hHMZ+\hUM-\UMB$. 

mRPA can be taken as one of the improvements of cRPA \cite{kotani_ab_2000,aryasetiawan_frequency-dependent_2004}
in the sense to determine screened Coulomb interaction without screening effects from the model space.
Until now, a variety of cRPA methods have been developed 
\cite{12s,13s,miyake_ab_2009,10s,9s,Wehling,sasioglu_effective_2011,11s,nomura2,14s,15s,biermann1,casula,misawa,16s,sasioglu_dftu_2014,shinaoka,swj-crpa,biermann2,hirayama}. For example, \c{S}a\c{s}{\i}o\v{g}lu, Freidlich and Bl\"uegel
\cite{sasioglu_effective_2011,sasioglu_dftu_2014} developed a convenient
cRPA method applicable to the case of entangled energy bands, 
while Miyake {\it et al.} \cite{miyake_ab_2009} treated the case in a different manner.
Nomura {\it et al.} showed a method to estimate the effective
interaction for impurity problems in DMFT 
\cite{nomura2}. Casula {\it et al.} showed a method beyond the RPA to include the band renormalization effects \cite{casula}.

In this paper, we apply mRPA to high-$T_c$
cuprate superconductors La$_2$CuO$_4$ ($T_c = 39$ K [\onlinecite{lsco-tc}], denoted by La) and
HgBa$_2$CuO$_4$ ($T_c = 98$ K [\onlinecite{hbco-tc}], denoted by Hg) to determine $\hHM$ of a two-orbital model \cite{sakakibara1,sakakibara2,sakakibarap,sakakibara3}.
After we determine $\hHM$,
we perform FLEX calculations to investigate superconductivity.
Our results are consistent with experiments.
Since this mRPA+FLEX procedure can be performed without parameters by hand, 
we can claim that relative height of $T_c$ among materials
 is evaluated just from crystal structures.
Thus, in principle, mRPA+FLEX can be used to 
find out a highest $T_c$ material among a lot of possible materials.

We like to emphasize importance of the two-orbital model \cite{sakakibara1,sakakibara2,sakakibarap,sakakibara3}.
Although the Fermi surface of cuprates consists of the $d_{x^2-y^2}$ orbital
mainly, Sakakibara {\it et al.} pointed out that 
hybridization of the $d_{x^2-y^2}$ orbital with the $d_{z^2}$ orbital 
\cite{kamimura,eto,freeman,wang,hozoi,uebelacker,hozoi-laad}
 is very important.
 This can be represented by the two-orbital model.
Sakakibara's FLEX calculation showed
that the hybridization degrades spin-fluctuation-mediated superconductivity.
This explains the difference of $T_c$ between La and Hg cuprates \cite{sakakibara1}.
A recent photoemission experiment for La cuprate
has captured significant orbital hybridization effects 
\cite{Matt2018}.

\begin{figure}[!t]
\includegraphics[width=7cm]{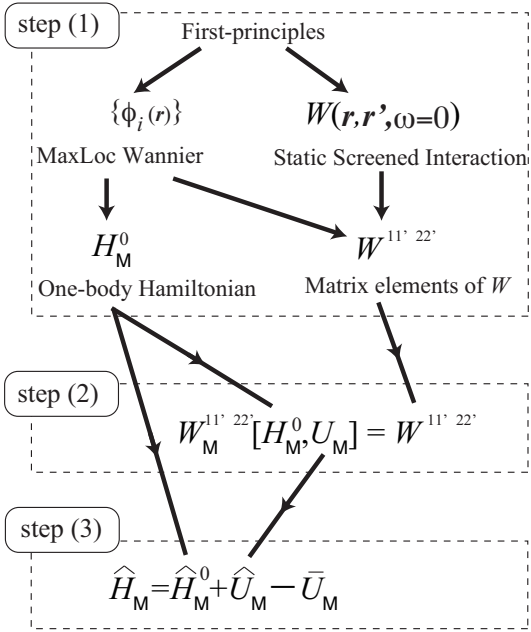}
\caption{How mRPA determines a model Hamiltonian $\hHM$.
Note that quantities with subscript {\sf M} are for the model Hamiltonian.
At step (1), we obtain 
one-body Hamiltonian $\HMZ$ and RPA screened Coulomb
interaction $\WFP^{11'22'}$ in a first-principles calculation.
At step (2), we obtain effective interaction $\UM$ in the model,
where we require $\WM^{11'22'}$ should be the same as $\WFP^{11'22'}$.
At step (3), we determine $\UMB$, which is to remove the double counting
in the one-body term.
}
\label{fig:mRPA} 
\end{figure}

\begin{figure*}[!ht]
\includegraphics[width=18cm]{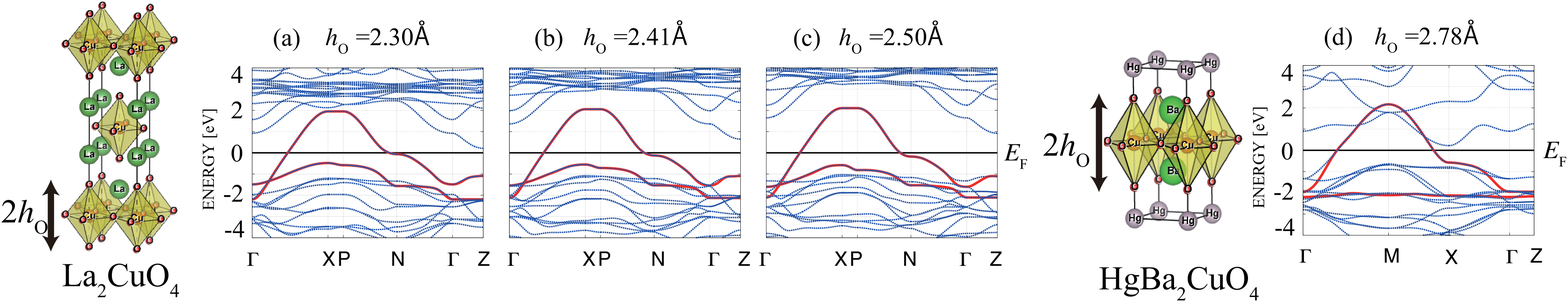}
\caption{(Color online) Crystal structures and band structures of La$_2$CuO$_4$(a)-(c) and HgBa$_2$CuO$_4$(d). 
Blue dashed lines are for the LDA band structures; red solid lines are for the two-orbital models.
The cases (a)-(c) are for varying the apical oxygen height $h_{\rm O}$. 
The cases (b) and (d) are with the experimental $h_{\rm O}$ \cite{lsco,hbco}.
\label{fig:band-st}
} 
\end{figure*}  

\section{method}\label{sec:method}
Let us summarize the formulation of mRPA in Ref. \onlinecite{mRPA}.
First of all, we have to parametrize the interaction $\hUM$ of the model Hamiltonian 
so that $\hUM$  is specified by finite numbers of parameters.
Fig. \ref{fig:mRPA} is a chart about how we determine $\hHM$.
Step (1) is by first-principles calculations, 
and step (2), (3) are by model calculations.
In this paper, we will treat the on-site-only interaction of the two-orbital model
specified by four parameters. 

In step (1) of Fig. \ref{fig:mRPA},
we first perform a self-consistent calculation in 
first-principles method. Then we can 
obtain one-body Hamiltonian $\hHMZ$ 
in the standard procedure of maximally localized Wannier function
\cite{marzari_maximally_1997,souza_maximally_2001}. 
In addition, we calculate static screened Coulomb interaction 
$W(\bfr,\bfr',\omega=0)$ in RPA.
Hereafter we omit $\omega=0$ since we treat only the static case in this paper.
Then we calculate matrix elements $\WFP^{11'22'}$ of the matrix $\WFP$, defined as
\begin{eqnarray}
&&\WFP^{11'22'}=  \left( 11' \right| W \left|22' \right)\nonumber\\
&&=\int d^3\bfr d^3\bfr' \ww^*_1(\bfr)\ww_{1'}(\bfr)  W(\bfr,\bfr')\ww^*_2(\bfr') \ww_{2'}(\bfr') , \label{eq:wfp}
\end{eqnarray}
where $\{\ww_1(\bfr)\}=\{\ww_{i_1\bm{R_1}}(\bfr)\}$ are the Wannier functions.
${\bm R}$ and $i$ denote a position of primitive cell and an orbital
in each cell, respectively. 
The number of elements $\WM^{11'22'}$
is the same as the number of elements $\UM ^{11'22'}$.
Calculations are performed with {\sf ecalj} package available from Git-hub \cite{ecalj}.

In step (2),
we determine $\UM$, so that it satisfies
\begin{eqnarray}
\WM^{11'22'}[\HMZ,\UM]=\WFP^{11'22'},
 \label{eq:mrpa}
\end{eqnarray}
where a functional $\WM^{11'22'}[\HMZ,\UM]$ is a screened 
interaction in RPA calculated from $\HMZ$ and $\UM$.
Here $\HMZ$ denotes the matrix whose elements are $H^{0,12}_{\sf M}$;
$\UM$ denotes the matrix whose elements are $\UM^{11'22'}$ as well.
$\hHMZ$ is the second quantized operator made of the matrix $\HMZ$,
$\hUM$ as well.
The functional is defined just in the model calculation; we do not treat
quantities spatially dependent on $\bfr$.
\req{eq:mrpa} is a key assumption of mRPA; we require that the screened
interaction in a model should be the same as
those of theoretical correspondence in the first-principles calculation.
 
Let us detail the functional $\WM^{11'22'}[\HMZ,\UM]$.
With non-interacting polarization function $\PM[\HMZ]$ of a model,
we have effective interaction $\WM$ in RPA as
\begin{eqnarray}
\WM[\HMZ,\UM]=\frac{1}{1-\UM\PM [\HMZ]} \UM.\label{eq:wmat}
\end{eqnarray}
Hereafter we omit $\HMZ$ in $\PM$ for simplicity.
Here we only treat non-magnetic case.
From \req{eq:wmat}, we have
\begin{eqnarray}
&&\WM^{i_1i_{1'}i_{2}i_{2'}}[\HMZ,\UM]=
 \frac{1}{N} \sum_{{\bm q}} \left[ \frac{1}{{\bm 1}-\UM\PM  ({\bm q})} \UM\right]_{i_1i_{1'}i_{2}i_{2'}}.  \label{eq:wm}
\end{eqnarray}
for on-site interactions $\UM$ and $\WM$.
\req{eq:wm} is used in \req{eq:mrpa} so as to determine $\UM$.

In step (3), we evaluate the one-body double counting term $\UMB$
contained in the total model Hamiltonian $\hHM$. It is written as
\begin{eqnarray}
\hHM= \widehat H_{\sf M}^{0} + \hUM - \UMB. 
\label{eq:model}
\end{eqnarray}
To determine $\UMB$, 
we require that the contribution from $\hUM$ and
that from $\UMB$ completely cancel
when we treat $\hUM$ in a mean-field approximation. 
The mean-field approximation should theoretically correspond to 
the first-principle method from which we start.
For example, if we use quasi-particle self-consistent $GW$ (QSGW) 
\cite{kotani_quasiparticle_2007,kotani_quasiparticle_2014,deguchi2016} as the first-principle method, we have to
use QSGW to treat the model of \req{eq:model}. Then $\UMB$
is made of the Hartree term and the static self-energy term in the
model. These terms cancel the effect of $\hUM$ when QSGW is applied to.
In this case, we have reasonable theoretical correspondence between the
first-principle calculation and model calculation.
However, if we use LDA as the first-principle method,
we have no corresponding mean-field approximation.
Thus we cannot uniquely determine $\UMB$.
Instead of determining $\UMB$, we use a practical method 
to avoid double counting in FLEX (see Sec. \ref{sec:flex}).

Let us recall the procedure of cRPA as a reference to mRPA.
The effective interaction of cRPA ($\Um$) is determined based on the requirement 
\begin{eqnarray}
\frac{1}{1- v P} v &=&\frac{1}{1- \Um\Pm} \Um,
\label{eq:crpaw}
\end{eqnarray}
where $v({\bm r},{\bm r'})$ is the bare Coulomb interaction,
$\Pm({\bm r},{\bm r'})$ is the polarization function within the 
model space spanned by the maximally localized Wannier functions.
\req{eq:crpaw} leads to 
\begin{eqnarray}
\Um &=&\frac{1}{1-  v(P-\Pm)}v \label{eq:crpa}.
\end{eqnarray}
Then we calculate the on-site matrix elements $\Um^{122'1'}=(11'| \Um|22')$.

Generally speaking, this cRPA procedures of \req{eq:crpa} 
cannot be applicable to systems with entangled energy bands if
the positive definiteness of $-(P-\Pm)$ in \req{eq:crpa} is not satisfied.
In fact, we have checked that $-(P-\Pm)$ 
do not satisfy the positive definiteness for La and Hg. 
Thus we need to use a modified $\Pm$ satisfying the positive definiteness in a
manner given  by \c{S}a\c{s}{\i}o\v{g}lu, Freidlich and Bl\"uegel \cite{sasioglu_effective_2011,sasioglu_dftu_2014}.
In their method, such $\Pm$ is given in Eq. (60) in Ref. \onlinecite{sasioglu_dftu_2014} as
\begin{eqnarray}
\Pm({\bm r},{\bm r'})
=\sum_{i}^{\rm occ}
 \sum_{j}^{\rm  unocc} 
\frac{ -2(c_i c_j)^2 \phi_{i}(\bfr) \phi^*_{j}(\bfr)
 \phi_{j}(\bfr') \phi^*_{i}(\bfr')}{\epsilon_j-\epsilon_i},
\label{eq:pm}
\end{eqnarray}
where $\phi_{i}$ is the eigenfunctions. The probability factor $c_i$ is 
the norm for $\phi_{i}(\bfr)$ projected into the model space spanned by the Wannier functions
(See Eq. (58) in Ref. \onlinecite{sasioglu_dftu_2014}). The composite index
$i=(\bfk,n)$ is for the wave number $\bfk$ and the band index $n$.
Apparently, $0\le c_{\bfk n} \le 1$ and $\sum_n (c_{\bfk n})^2=1$ are satisfied for given $\bfk$.
Thus $-(P-\Pm)$ is clearly positive definite because it is calculated just
from the equation with $1-(c_i c_j)^2$ instead of $-(c_i c_j)^2$ in the numerator of \req{eq:pm}. 

\begin{table}[!h]
\caption{The interactions of mRPA ($\UM$) and cRPA ($\Um$) 
in a three-orbital model for SrVO$_3$, where $d_{xy},d_{yz}$, and $d_{zx}$
orbitals are considered.
$U$, $U'$, $J$ are the intra-orbital, inter-orbital, and exchange
interactions, respectively.
The static screened interaction $W$ is also shown
in the same manner as $\UM$.}
\label{tab:svo}
\begin{tabular}{l c c | c c} \Hline
  SrVO$_3$ &  & mRPA   & cRPA       \\
 \Hline 
 [eV]& $\WFP$ & \hspace{0.2cm}$\UM$ \hspace{0.2cm} &   \hspace{0.2cm} $\Um$
	     \hspace{0.2cm} & \hspace{0.2cm}  \\\hline 
$U$    &  0.852 & 2.82  &	 3.12  \\
$U'$        & 0.248  & 1.88  &	2.17  \\
$J$  & 0.290 & 0.442 & 0.448	  \\
\Hline
\end{tabular}
\end{table}

As a check for our implementation of mRPA and cRPA, we show $\Um$ and $\UM$
for SrVO$_3$ where three 3$d$ bands spanning model space are clearly separated
from the other bands. 
In this case, we can expect that non-zero $c_i$ are not widely distributed
among energy bands. Only $c_i$ for the three 3$d$ bands are almost
unity, while others are almost zero. In this case, as shown in Table
\ref{tab:svo}, $\Um$ is close to $\UM$: $U$ of $\UM$, 2.82 eV, is only a little smaller than
$U$ of $\Um$, 3.12 eV. This is reasonable since both mRPA and cRPA are to remove screening 
effect related to the model space, although we treat only the on-site
interactions in mRPA. 
The difference $2.82-3.12=-0.30$ eV may be mainly explained by the effect
of off-site interactions. 
To check this, we apply mRPA using Eq. (9) of Ref. \onlinecite{mRPA} 
including the interactions between all vanadium sites.
In this case, the values obtained in mRPA should be in agreement with that of cRPA in principle.
We find that $U$ of $\UM$ become larger 
\footnote{In our previous paper \cite{mRPA}, we made a wrong statement that $\UM$ 
would become smaller if we consider off-site interactions.}
 to be 3.33 eV, slightly
overshoots but becomes closer to 3.12 eV.
Still remaining difference $3.33-3.12=0.21$ eV may be due to
 detailed differences of formalisms and numerical treatment.

\section{Result for effective interaction}
Following the chart of Fig. \ref{fig:mRPA}, 
we apply mRPA to single-layered cuprates, La and Hg, 
to obtain the two-orbital Hubbard model \cite{sakakibara1}, where we start from LDA calculations.
We show their experimental crystal structures \cite{lsco,hbco} in Fig. \ref{fig:band-st}, together with
their LDA band structures in (b) and (d), where we superpose the energy bands of the two-orbital models.
In addition, we treat hypothetical cases varying apical oxygen height
$h_{\rm O}$ in La, (a) and (c), in order to clarify differences between mRPA and cRPA.
Here $h_{\rm O}$ is defined as the distance shown in Fig. \ref{fig:band-st}.
The matrix $\UM$ of the two-orbital model is represented as
\begin{equation}
\UM = \begin{pmatrix} U^{x^2-y^2} && 0 && 0 && U'\\
0 && U^{J'} && U^J && 0\\
0 && U^J && U^{J'} && 0\\
U' && 0 && 0 && U^{z^2}
 \end{pmatrix}, \label{eq:matU}
 \end{equation}
where the indices of the matrix $\UM$ takes $d_{x^2-y^2}d_{x^2-y^2}$,
$d_{x^2-y^2}d_{z^2}$, $d_{z^2}d_{x^2-y^2}$, and $d_{z^2}d_{z^2}$.
Here $U'$ is inter-orbital Coulomb interactions; $U^J=U^{J'}$ are exchange interactions.
Other interactions such as $\WM$ are represented as well.

\begin{table}[!h]
\caption{The interactions of mRPA ($\UM$) and cRPA ($\Um$) 
for the experimentally observed crystal structure of La$_2$CuO$_4$ and HgBa$_2$CuO$_4$ \cite{lsco,hbco}.
The elements of $\WFP$ are defined in the same manner as $\UM$ (see text).
}
\label{tab:mRPA}
\begin{tabular}{l c c | c c} \Hline
La$_2$CuO$_4$&     & mRPA & cRPA    \\
 \Hline 
 [eV]& $\WFP$ & \hspace{0.2cm}$\UM$ \hspace{0.2cm} &   \hspace{0.2cm} $\Um$  \hspace{0.2cm} & \hspace{0.2cm}  \\\hline 
$U^{x^2-y^2}$    & 0.747& 2.76  &	3.14  \\
$U^{z^2}$        & 1.58 & 2.63	&	2.95  \\
$U'$	         & 0.370& 1.64	&	2.01  \\
$U^J$	         & 0.273& 0.44 	&	0.41  \\
\Hline
HgBa$_2$CuO$_4$&     & mRPA  & cRPA&      \\
 \Hline 
 [eV]& $\WFP$ & \hspace{0.2cm}$\UM$ \hspace{0.2cm} &   \hspace{0.2cm} $\Um$  \hspace{0.2cm} &\hspace{0.2cm}  \\\hline 
$U^{x^2-y^2}$ & 0.820&  2.99  &	2.14	& \\
$U^{z^2}$     & 3.83 &	5.47  &	4.93 	& \\
$U'$	      & 0.724&	2.62  &	1.92	& \\
 $U^J$	      & 0.460&	0.67  &	0.58 	& \\
\Hline
\end{tabular}
\end{table}

In Table \ref{tab:mRPA}, we show values of $\UM$ for La
and Hg (Fig. \ref{fig:band-st}(b) and \ref{fig:band-st}(d)), together with values of $\WFP$
\footnote{We use the tetrahedron method \cite{kotani_quasiparticle_2007,kotani_quasiparticle_2014} 
 in the Brillouin zone to calculate the matrix $W$ and $U_{\sf m}$, where we use
 $8\times 8 \times 8 (8\times 8 \times 4)$ $k$-points for La$_2$CuO$_4$(HgBa$_2$CuO$_4$).
 For a model calculation to determine $U^{122'1'}_{\sf M}$, we take $64\times 64 \times 4$ $k$-point grids
 for discrete summation. We use dense enough 4096 Matsubara meshes at
 $T=0.005$ eV (virtually equal to $T=0$ eV.)}.
At first, let us compare $\WFP$ for La and Hg.
We see a little difference on $\WFP ^{x^2-y^2}$ (0.747eV vs. 0.820 eV),
while larger difference on $\WFP ^{z^2}$ (1.58 eV vs. 3.83 eV). 
This is expected since Hg is more anisotropic than La, as indicated by the size of $h_{\rm O}$.
From these $\WFP$ and the band structure of the two-orbital model, we have obtained
$\UM$ shown in Table \ref{tab:mRPA}. 
We see that ratios $\UM/\WFP$ are similar for La and Hg, that is,
$2.76/0.747 \sim 2.99/0.820$ for $W^{x^2-y^2}$, other elements as well.
This is consistent with the similarity of the band structure shown in Fig.\ref{fig:band-st} (b) and (d).

We find that $\UM^{x^2-y^2}$ is roughly estimated by
\begin{eqnarray}
\UM^{x^2-y^2} \sim \frac{\WFP^{x^2-y^2}}{1+\WFP^{x^2-y^2} \PM^{x^2-y^2}},
\label{eq:umapprox}
\end{eqnarray}
where $\PM^{x^2-y^2}$ is the diagonal elements of the Brillouin zone average of
$\PM(\bm q)$. \req{eq:umapprox} is derived from \req{eq:wm} by
replacing $\PM(\bm q)$ with the average.
Let us evaluate \req{eq:umapprox}.
Our calculation gives $\PM^{x^2-y^2}=-0.97$ eV$^{-1}$ for La, $-0.91$
eV$^{-1}$ for Hg. The little difference $-0.06=(-0.97)-(-0.91)$ eV$^{-1}$
corresponds to the little difference of the band structures of the two-orbital models shown in Fig. \ref{fig:band-st}(b) and \ref{fig:band-st}(d).
Together with the values of $\WFP^{x^2-y^2}=0.747,0.820$ eV in Table \ref{tab:mRPA}, 
\req{eq:umapprox} gives $\UM^{x^2-y^2} \sim 2.71$ eV for La and $\sim 3.23$ eV
for Hg. These are roughly in agreements with $\UM^{x^2-y^2}=$ 2.76 and 2.99 eV in Table \ref{tab:mRPA}.
This analysis indicates that the difference of $\UM^{x^2-y^2}$ between La
and Hg is mainly due to the difference of $\WFP^{x^2-y^2}$.

In Table \ref{tab:mRPA}, we also show cRPA values $\Um$ for comparison.
For La, Table \ref{tab:mRPA} shows that $\Um$ gives good agreement with $\UM$, 
a little smaller as in the case of SrVO$_3$ in Table \ref{tab:svo}.
On the other hand, we see large discrepancy for Hg :
$\Um^{x^2-y^2}=$2.14 eV is much smaller than $\UM^{x^2-y^2}=$2.99 eV.
This difference can be explained by \req{eq:pm} with factors $c_i$.
In Hg, we see a stronger $d$-$p$ hybridization in Fig.\ref{fig:band-st} (d) than
La; the position of Cu-$d_{x^2-y^2}$ band is pushed down to be in the
middle of oxygen bands. This means that non-zero $c_i$ are 
more distributed among the oxygen bands in the case of Hg 
than in the case of La. This can be a reason to make the effective size 
of $\Pm$ smaller than $\PM$ in the case of Hg, 
resulting the smaller $\Um$.

To confirm the effect of hybridization, we calculate $\Um$ and $\UM$
by varying $h_{\rm O}$ for La. As discussed in Ref. \onlinecite{sakakibara1},
 $h_{\rm O}$ is a key quantity to determine the critical temperatures 
 of superconductors \cite{maekawa,Andersen,pavarini,mori,weber1,weber2}.
We can see $h_{\rm O}$ works as a control parameter of hybridization \cite{swj-crpa,weber1,weber2}.
That is, as shown in \ref{fig:band-st}(a)-(c), higher $h_{\rm O}$ pushes
down Cu-$d_{x^2-y^2}$ levels more, resulting larger hybridization with
oxygen bands. 
Fig. \ref{fig:band-st}(d) for Hg can be taken as a case with highest $h_{\rm O}$.

\begin{figure}[!t]
\includegraphics[width=9cm]{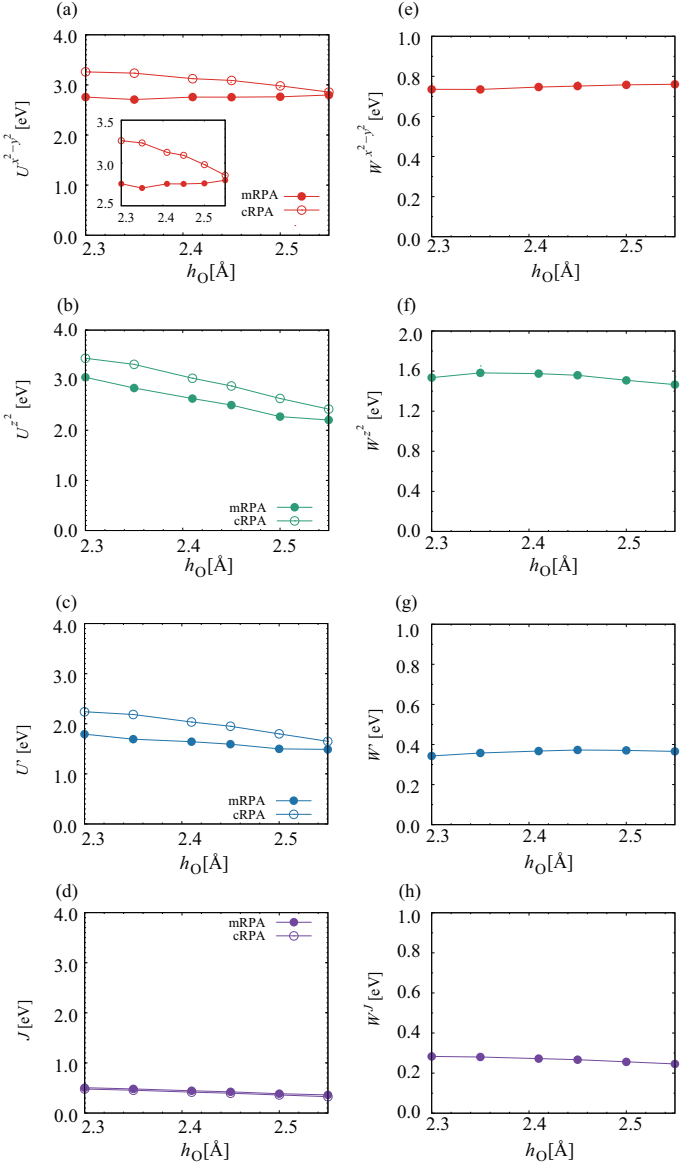}
\caption{(Color online) The elements of $\UM$(mRPA), $\Um$(cRPA), and $\WFP$ are plotted as a
function of $h_{\rm O}$. 
Details of numerical settings are shown in the text.
Note that $\WM^{11'22'}[\UM]=\WFP^{11'22'}$ is satisfied at any values of $h_{\rm O}$.
Panels (a) and (e) indicate that $U^{x^2-y^2}$ for cRPA is affected by the localization of 
Wannier functions (see text). 
\label{fig:fig-plot}
} 
\end{figure}

In Fig. \ref{fig:fig-plot}, we plot $\UM$ and $\Um$ together with $\WFP$. 
Let us focus on Fig. \ref{fig:fig-plot}(a) and (e).
As a function of $h_{\rm O}$, $\WFP^{x^2-y^2}$ is almost constant.
In addition, the energy bands of the two-orbital model
change little as shown in Fig. \ref{fig:band-st}(a)-(c).
Thus it is reasonable that $\UM^{x^2-y^2}$ changes 
little in Fig. \ref{fig:fig-plot}(a), because of \req{eq:umapprox}.
On the other hand, $\Um^{x^2-y^2}$ decreases rapidly 
when $h_{\rm O}$ becomes higher. 
This means that $\Pm$ becomes smaller for higher $h_{\rm O}$.
As in the case of Hg case, we think this is because of larger
hybridization of Cu-$d_{x^2-y^2}$ bands with oxygen bands.

Our mRPA and cRPA results are rather different.
In Ref. \onlinecite{swj-crpa}, we treated a variety of layered
cuprates, where we show that the effective interaction for La is larger
than that for Hg as shown by $\Um$ in Table \ref{tab:mRPA},
based on the cRPA calculations. In addition, we showed the effective
interactions are
controlled by $h_{\rm O}$ as shown in $\Um$ in Fig. \ref{fig:fig-plot}.
Even though we do  not need to modify the overall conclusion in
Ref. \onlinecite{swj-crpa},
we should not take such effective interactions as suitable for
Hubbard models.
Along the logic of mRPA, we should use $\UM$ instead of $\Um$.

\section{FLEX calculation for superconductivity}\label{sec:flex}
For the model Hamiltonian $\hHM$ obtained from mRPA,
we perform two-orbital FLEX calculations
to obtain dressed Green's functions $G_{ij}(k)$ \cite{FLEX,lichtenstein_ab_1998,mFLEX1,mFLEX2,takimoto}.
Here $k=({\bm k},i\omega_n)$ is a composite index made of the wave vector
$\bm k$ and the Matsubara frequency $i\omega_n$. 
The band index $i$ takes 1 or 2.
We calculate only the optimally doped case for $T_c$ (15\% doping).
We take $32\times 32\times 4$ $k$-meshes and 1024 Matsubara frequencies.

Let us remind step (3) in Fig. \ref{fig:mRPA} to determine the counter one-body
term $\UMB$. Instead of LDA, let us consider QSGW case first. Theoretically, it is
easier since QSGW is a method directly applicable even to a model
Hamiltonian, where QSGW determines a mean-field one-body Hamiltonian for
the model. We first determine $\hHMZ$ in QSGW by the first-principle QSGW calculation and
the Wannier function method in the step (1) of mRPA.
Then we can determine $\hUM$ in the step (2) of mRPA.
In the step (3), we apply the QSGW method to the model Hamiltonian
$\hHM = \hHMZ + \hUM - \UMB$, where yet unknown term $\UMB$ is included. 
Here $\UMB$ is determined so that the QSGW applied to $\HM$ do give
the mean-field one-body Hamiltonian $\hHMZ$. That is, the effect of $\hUM$
to the one-body Hamiltonian is completely canceled by $\UMB$.

When we start from LDA instead of QSGW, we have no unique way to
determine $\UMB$ since LDA cannot be applicable to the model Hamiltonian.
Thus we need some assumption to follow the case of QSGW.
Here we identify the static part of the self-energy 
$\Sigma(\bm{k},0)$ as $\UMB$ 
(our definition of $\Sigma(\bm{k},0)$ here includes the Hartree term).
In other words, if we perform a static FLEX calculation only with
$\Sigma(\bm{k},0)$, we reproduce the one-body Hamiltonian of LDA.
This method is equivalent to Eq. (5) in Ref. \onlinecite{ikeda}.
We simply assume FLEX is not for the mean-field part, 
but for the $\omega-$dependent self-energy part.

Here we investigate superconductivity in the two-orbital model.
By substituting $G_{ij}(k)$ 
into the linearized Eliashberg equation,
\begin{eqnarray}
\lambda \Delta_{ij}(k)&=&-\frac{T}{N}
\sum_{q,m_i}  V_{im_1m_4j} (q) G_{m_1m_2} (k-q)\nonumber\\
&&\times \Delta_{m_2m_3}(k-q) G_{m_4m_3}(-k+q),\label{eq:elia}
\end{eqnarray}
we obtain the gap function $\Delta_{ij}(k)$ as an eigenstate and its eigenvalue $\lambda$,
where $V(q)$ is the singlet pairing interaction as described in Eq. (2)-(7) of Ref. \onlinecite{sakakibara2}.
The largest $\lambda$ reaches unity at $T=T_c$.
Since $\lambda$ is monotonic and increasing function of $T^{-1}$,
we use $\lambda$ at  $T=0.01$ eV as a qualitative measure of $T_c$ 
instead of calculating at $T_c$.
In some FLEX calculations, $\lambda$ at fixed temperature is used to compare 
relative height of $T_c$ among similar materials \cite{PRB79,ikeda}.
We obtain $\lambda= 0.50$ for La and 0.71 for Hg.
This is qualitatively consistent with the experimental observation
that Hg ($T_c = 98$ K) is higher than La ($T_c = 39$ K) \cite{lsco-tc,hbco-tc}.

\begin{figure}[!t]
\includegraphics[width=6cm]{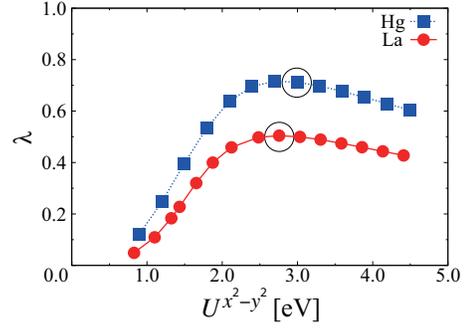}
\vspace{20pt}
\caption{(Color online) The eigenvalues $\lambda$ of the Eliashberg equation 
 are plotted as a function of $U^{x^2-y^2}$. 
 Here the temperature is 0.01 eV.
Red filled circles show the value for La and blue squares for Hg.
Open circles indicate the results obtained with the value shown in table \ref{tab:mRPA}.
}
\label{fig:flex}
\end{figure}

To investigate how $\UM$ affects $\lambda$ in more detail,
we perform calculations by rescaling $\UM$ hypothetically.
We plot $\lambda$ as a function of $U^{x^2-y^2}$ in Fig. \ref{fig:flex}.
In the calculation, $\hHMZ$ and the ratio between all the elements of
$\UM$ are fixed. 
We see that $\lambda$ increases rapidly with smaller $U^{x^2-y^2}$
and plateaus with larger $U^{x^2-y^2}$ in both materials.
The cases of original $U^{x^2-y^2}$ as shown in table \ref{tab:mRPA} are shown by open circles.
These are in the plateau region \footnote{The correlation between $U/t$ and $T_c$ is discussed 
 with Hubbard model calculations, e.g., in Ref. \onlinecite{yokoyama}.}.
Because of the small changes in the region,
$\lambda$ of the two cuprates do not change so much even if we use $\Um$ instead of $\UM$,
where $\lambda_{\sf cRPA}^{\sf La}=0.52$ and $\lambda_{\sf cRPA}^{\sf Hg}=0.64$.
The difference between La and Hg is mainly 
from the hybridization of the $d_{x^2-y^2}$ orbital with the $d_{z^2}$ orbital.
This is already examined by previous FLEX calculations with empirically determined interaction parameters \cite{sakakibara1}.
Sakakibara {\it et al}. already showed that FLEX reproduces the experimental trends of $T_c$  
(see Fig. 1(a) of Ref. \onlinecite{sakakibara3}).  
The detailed mechanism how the hybridization affects $T_c$
was discussed in Sec. III D of Ref. \onlinecite{sakakibara2}.

\section{summary}
With mRPA, we obtain the two-orbital Hubbard models for
La$_2$CuO$_4$ and HgBa$_2$CuO$_4$ in first-principles. The main part of mRPA is how to determine the
on-site interaction parametrized by four parameters.
We see that the interactions are close to those in cRPA.
However, we see some differences.
A difference comes from the fact 
that the effective size of the polarization function $\Pm$ in cRPA becomes smaller
than $\PM$ in mRPA.
This is because that the probability factors $c_i$ in \req{eq:pm} 
are distributed among the oxygen bands when $d$-$p$ hybridization is strong, 
as in HgBa$_2$CuO$_4$.

For the models, we perform FLEX to evaluate superconductivity.
The results are consistent with experiments. 
With the interaction obtained in mRPA, we confirm that 
$T_c$ is not so strongly dependent on the scale of interaction.
Along the line of the combination of mRPA and FLEX, we will be able to predict new superconductors.

We appreciate discussions with Drs. Friedlich, \c{S}a\c{s}{\i}o\v{g}lu, Imada, Arita, Hirayama, Kuroki,
S. W. Jang, and M. J. Han. 
H.S. appreciates fruitful discussions with Drs. Misawa, Nomura, and Shinaoka.
This work was supported by JSPS KAKENHI (Grant No. 16K21175, 17K05499).
The computing resource is
supported by Computing System for Research in Kyushu University (ITO system),
the supercomputer system in RIKEN (HOKUSAI),
and the supercomputer system in ISSP (sekirei). 

\bibliography{refsk}

\end{document}